\RequirePackage{fix-cm}
\documentclass[aps,pra,reprint,superscriptaddress]{revtex4-2}

\RequirePackage{graphicx}

\usepackage{graphicx}
\usepackage{dcolumn}
\usepackage{bm}
\usepackage{lipsum}

\usepackage{hyperref, url}

\usepackage{amsmath}

\usepackage{tikz}

\usepackage{ctable}

\usetikzlibrary{arrows.meta}
\usepackage{pgfplots}
\usepgfplotslibrary{colormaps}
\pgfplotsset{compat=1.8}
\usetikzlibrary{pgfplots.groupplots,
                pgfplots.fillbetween}

\usetikzlibrary{decorations.pathreplacing}

\usepackage{xspace}
\newcommand{\pentatrap}{\textsc{Pentatrap}\xspace}

\begin{document}

\title{High-precision mass measurement of doubly magic $^{208}$Pb}
\author{Kathrin Kromer}
\email[]{kromer@mpi-hd.mpg.de}
\affiliation{Max-Planck-Institut für Kernphysik, 69117 Heidelberg, Germany}
\author{Chunhai Lyu}
\affiliation{Max-Planck-Institut für Kernphysik, 69117 Heidelberg, Germany}
\author{Menno Door}
\affiliation{Max-Planck-Institut für Kernphysik, 69117 Heidelberg, Germany}
\author{Pavel Filianin}
\affiliation{Max-Planck-Institut für Kernphysik, 69117 Heidelberg, Germany}
\author{Zoltán Harman}
\affiliation{Max-Planck-Institut für Kernphysik, 69117 Heidelberg, Germany}
\author{Jost Herkenhoff}
\affiliation{Max-Planck-Institut für Kernphysik, 69117 Heidelberg, Germany}
\author{Wenjia Huang}
\affiliation{Advanced Energy Science and Technology Guangdong Laboratory, Huizhou 516007, China}
\author{Christoph H. Keitel}
\affiliation{Max-Planck-Institut für Kernphysik, 69117 Heidelberg, Germany}
\author{Daniel Lange}
\affiliation{Max-Planck-Institut für Kernphysik, 69117 Heidelberg, Germany}
\affiliation{Ruprecht-Karls-Universität Heidelberg, 69117 Heidelberg, Germany}
\author{Yuri N. Novikov}
\affiliation{Department of Physics, St Petersburg State University, St Petersburg 198504, Russia}
\affiliation{NRC “Kurchatov Institute”-Petersburg Nuclear Physics Institute, Gatchina 188300, Russia}
\author{Christoph Schweiger}
\affiliation{Max-Planck-Institut für Kernphysik, 69117 Heidelberg, Germany}
\author{Sergey Eliseev}
\affiliation{Max-Planck-Institut für Kernphysik, 69117 Heidelberg, Germany}
\author{Klaus Blaum}
\affiliation{Max-Planck-Institut für Kernphysik, 69117 Heidelberg, Germany}

\begin{abstract} 
The absolute atomic mass of $^{208}$Pb has been determined with a fractional uncertainty of $7\times 10^{-11}$ by measuring the cyclotron-frequency ratio $R$ of $^{208}$Pb$^{41+}$ to $^{132}$Xe$^{26+}$ with the high-precision Penning-trap mass spectrometer \pentatrap and computing the binding energies $E_{\text{Pb}}$ and $E_{\text{Xe}}$ of the missing 41 and 26~atomic electrons, respectively, with the \textit{ab initio} fully relativistic multi-configuration Dirac-Hartree-Fock (MCDHF) method. $R$ has been measured with a relative precision of \hbox{$9\times 10^{-12}$}. $E_{\text{Pb}}$ and $E_{\text{Xe}}$ have been computed with an uncertainty of 9.1~eV and 2.1~eV, respectively, yielding $207.976\,650\,571(14)$~u ($\text{u}=9.314\,941\,024\,2(28)\times10^{8}$~eV/c$^2$) for the $^{208}$Pb neutral atomic mass. This result agrees within $1.2\sigma$ with that from the \textit{Atomic-Mass Evaluation} (AME) 2020, while improving the precision by almost two orders of magnitude. The new mass value directly improves the mass precision of 14 nuclides in the region of $Z=81-84$ and is the most precise mass value with $A>200$. Thus, the measurement establishes a new region of reference mass values which can be used e.g. for precision mass determination of transuranium nuclides, including the superheavies.


\end{abstract}

\maketitle

\section{Introduction}
Heavy and superheavy nuclides beyond the doubly magic nucleus of $^{208}$Pb can only exist due to nuclear shell effects holding them together by counteracting the rapidly increasing Coulomb repulsion with growing proton number $Z$~\cite{oganessianSynthesisIsotopesElements2006}. Insight into these quantum-mechanical nuclear structure effects can be derived from the masses of such nuclides. In addition to some direct heavy mass measurements~\cite{blockDirectMassMeasurements2010,dworschakPenningTrapMass2010,ramirezDirectMappingNuclear2012,eibachDirectHighprecisionMass2014}, a network of nuclear transitions and relative mass measurements, i.e. the \textit{Atomic-Mass Evaluation} (AME), provides mass values for most heavy and superheavy nuclides by tracing them back to a few well-known masses of uranium isotopes~\cite{wangAME2020Atomic2021}. However, no nuclide beyond $Z = 70$ can be found whose mass is known to a relative precision of better than $2\times 10^{-9}$ to act as a precise reference point for these heavy elements. This directly limits the achievable precision in the heavier mass regions and can possibly lead to tensions or shifts of the relative measured masses due to their referencing to only one reference point. The limitations by mass dependent shifts can be reduced significantly once there is a reference mass with similar mass known to high precision~\cite{kellerbauerDirectAbsoluteMass2003}. The need for new anchor points for the AME arose during recent mass measurements with TRIGA-TRAP~\cite{eibachDirectHighprecisionMass2014,ketelaerTRIGASPECSetupMass2008} at the research reactor TRIGA in Mainz, specifically, an improved absolute mass of $^{208}$Pb~\cite{grundFirstOnlineOperation2020}. Measuring this mass will also directly improve the masses of several Pb isotopes and other nuclides in this mass region~\cite{wangAME2020Atomic2021}. 

In addition to the impact as a mass reference for other mass measurements, the mass of $^{208}$Pb will soon be needed when the magnetic moment, or the $g$-factor, of the bound electron of hydrogen-like $^{208}$Pb is planned to be determined by the Penning-trap experiments Alphatrap at the MPIK in Heidelberg~\cite{sturmHighprecisionMeasurementAtomic2014} and Artemis at GSI Darmstadt~\cite{vogelElectronMagneticMoment2019}. This measurement could be the most stringent test of bound-state quantum electrodynamics in strong fields. The error of the mass of the nucleus, however, enters the error budget and therefore needs to be known to high precision~\cite{kohlerIsotopeDependenceZeeman2016}. With the results of this paper, the error of the mass of $^{208}$Pb will be negligible in future $g$-factor determinations. 

Based on the accurate absolute mass of $^{132}$Xe~\cite{redshaw2009improved,rischkaMassDifferenceMeasurementsHeavy2020}, in this paper, we present a determination of the absolute atomic mass of $^{208}$Pb with a fractional uncertainty of $7\times 10^{-11}$. This is the result of measuring the cyclotron-frequency ratio of $^{208}$Pb$^{41+}$ and $^{132}$Xe$^{26+}$ with the high-precision Penning-trap mass spectrometer \pentatrap~\cite{reppPENTATRAPNovelCryogenic2012,rouxTrapDesignPENTATRAP2012} in combination with a computation of the binding energies of the missing 41 and 26~atomic electrons, respectively, using the \textit{ab initio} fully relativistic multi-configuration Dirac-Hartree-Fock (MCDHF) method. The masses of $^{132}$Xe$^{26+}$ and $^{208}$Pb$^{41+}$ can be related to their neutral counterparts via 
\begin{eqnarray}
m\left({}^{132}\text{Xe}^{26+}\right)&=&m\left({}^{132}\text{Xe}\right)-26m_e+E_{\text{Xe}},\\
m\left({}^{208}\text{Pb}^{41+}\right)&=&m\left({}^{208}\text{Pb}\right)-41m_e+E_{\text{Pb}},
\end{eqnarray} 
with $m_e=5.485\,799\,090\,65(16)\times10^{-4}$~u being the electron rest mass~\cite{zatorskiExtractionElectronMass2017} and $m\left({}^{132}\text{Xe}\right)=131.904\,155\,086(10)$~u being the mass of a neutral ${}^{132}$Xe atom~\cite{redshaw2009improved,rischkaMassDifferenceMeasurementsHeavy2020}, each has a relative accuracy of $2.9\times10^{-11}$ and $7.6\times10^{-11}$, respectively. $E_{\text{Xe}}$ and $E_{\text{Pb}}$ are the binding-energy differences that represent the energies required to ionize the outermost 26 and 41 electrons, respectively, from neutral Xe and Pb atoms. With the mass ratio 
\begin{eqnarray}
R&=&\frac{m\left({}^{208}\text{Pb}^{41+}\right)}{m\left({}^{132}\text{Xe}^{26+}\right)}
\end{eqnarray} 
being experimentally measured, one can improve the accuracy of the absolute mass of $^{208}$Pb via 
\begin{eqnarray}
&&m\left({}^{208}\text{Pb}\right)\nonumber\\
&=&R\left[m\left({}^{132}\text{Xe}\right)+26m_e-E_{\text{Xe}}\right]+41m_e-E_{\text{Pb}} \text{ ,}
\end{eqnarray} 
based on the theoretically calculated $E_{\text{Xe}}$ and $E_{\text{Pb}}$. By improving the mass of $^{208}$Pb the masses of other Pb isotopes and nearby elements can be improved accordingly since they are linked via decays of which the energy has been measured. 

\section{Experimental and theoretical methods}
\begin{figure*}
\centering
\includegraphics[width=17.0cm]{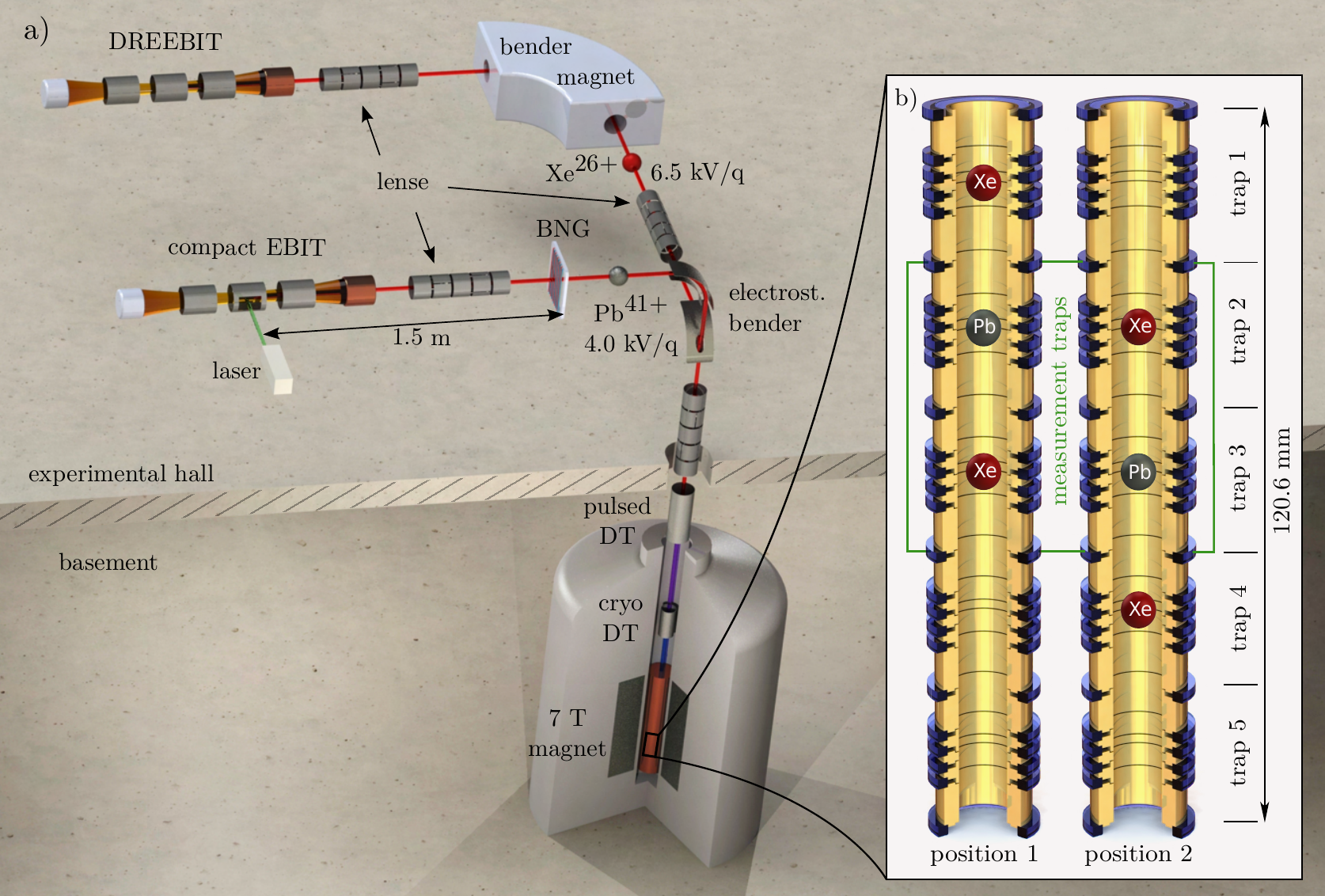}
  \caption{a) Schematic illustration of the ion production section, the two beamlines, and the combined deceleration region ending in the trap chamber. The ion bunches with their respective energies are shown after they have been $q/m$ selected by the bender magnet or the Bradbury-Nielson gate (BNG). b) Schematic drawing of the Penning-trap tower with two measurement ion configurations. The ions are moved from position 1 to 2 or back every $\approx 15$~min. The frequency measurements are carried out in traps 2 and 3. Traps 1 and 4 are utilized as storage traps and trap 5 is currently not in use but is planned to be used to monitor magnetic field fluctuations in the future. }
  \label{fig:Beamlines}
\end{figure*}

If one introduces a charged particle into a magnetic field $B$, it will describe a free space cyclotron motion with the frequency $\omega_c = \frac{q}{m}B$, with $q/m$ being the charge-to-mass ratio. The working principle of a Penning trap is based on a strong homogeneous magnetic field in combination with an electrostatic quadrupole potential. While the electrostatic potential prevents the ion from escaping in axial direction, forcing it onto an oscillatory axial motion with frequency $\omega_z$, the magnetic field forces the ion in radial direction onto a circular orbit with a modified cyclotron frequency $\omega_+$. The cross product of the two fields in the Lorentz equation leads to an additional slow drift around the trap center called magnetron motion with frequency $\omega_-$. When comparing these three Penning-trap eigenfrequencies to the movement of a free charged particle in a purely magnetic field, it holds~\cite{brownPrecisionSpectroscopyCharged1982}: 
\begin{equation}
    \omega_c = \sqrt{\omega_+^2 +\omega_z^2 + \omega_-^2} \text{ }.
     \label{eq:invariance}
\end{equation}
From this equation we can see that the determination of eigenfrequencies of an ion in a Penning trap can be used to determine its mass, if the magnetic field inside the trap is known. However, a determination of a magnetic field of $ B \approx 7$~T inside a volume of just a few $10~\mu$m$^3$ to sufficient precision is not possible. Therefore, a relative measurement is chosen at \pentatrap, using a reference ion and a sequential measurement scheme to determine mass ratios~\cite{reppPENTATRAPNovelCryogenic2012}. Highly charged ions are used due to the advantage that with higher $q/m$ the modified cyclotron frequency increases and can therefore be measured to a higher relative precision. For each mass determination a reference nuclide and charge states have to be chosen that form a $q/m$ doublet with the nuclide of interest in order to largely suppress systematic effects in the cyclotron-frequency ratio determination~\cite{reppPENTATRAPNovelCryogenic2012,rouxTrapDesignPENTATRAP2012}. The advantage being, that with $q/m$ doublets the same trapping voltage can be used to match the axial frequency to the detection tank circuit's resonance frequency. Using the same trapping voltage reduces systematic shifts due to trap anharmonicities. In addition, the absolute mass of the reference nuclide has to be known better than the aimed uncertainty of the mass of the nuclide of interest. More technical restrictions are posed by the production of the reference ion, limited by binding energies and the availability of probe material. For these reasons, the near $q/m$ doublet $^{208}$Pb$^{41+}$ ($q/m= 0.197\,138~e/\text{u}$) and $^{132}$Xe$^{26+}$ ($q/m= 0.197\,113~e/\text{u}$)~\cite{redshaw2009improved,rischkaMassDifferenceMeasurementsHeavy2020} was chosen. The $^{132}$Xe$^{26+}$ ion was created from a gaseous natural source inside a commercial \textit{Dresden electron beam ion trap} (DREEBIT)~\cite{ovsyannikovFirstInvestigationsWarm1999,StartseiteDREEBITGmbH}. The DREEBIT is connected to a beamline with a large bender magnet for $q/m$ selection, see \autoref{fig:Beamlines}a) upper beamline. The $^{208}$Pb$^{41+}$ ion was produced in a Heidelberg Compact electron beam ion trap (compact EBIT) ~\cite{mickeHeidelbergCompactElectron2018} equipped with an in-trap laser-desorption target of monoisotopic $^{208}$Pb~\cite{schweigerProductionHighlyCharged2019}. After ion breeding, the $q/m$ selection was achieved using the time-of-flight separation technique with fast high-voltage switches recently developed at the MPIK ~\cite{Schweiger2021}, supplying the voltages to a Bradbury-Nielson gate~\cite{bradburyAbsoluteValuesElectron1936}, see \autoref{fig:Beamlines}a) lower beamline. 
Once the ions were selected and decelerated by two pulsed drift tubes, they were consecutively trapped in the first of \pentatrap's five traps and transported down to their individual traps.

Due to the five stacked Penning traps available, see \autoref{fig:Beamlines}b), a simultaneous measurement in two traps is possible, increasing the measurement speed by higher statistics and offering up the opportunity for cross checks between the traps and several analysis methods. Out of the other three traps, two are needed for ion storage and one trap is planned for monitoring, however, currently not in use. 

The ion's frequencies depend on the magnetic field and the electrostatic potential. All environmental influences on these quantities need to be stabilized over the duration of the measurement. For this, the \pentatrap laboratory is temperature-stabilized to \hbox{$\delta T <50$~mK/h} and the height of the liquid helium level $z_{\text{lHe}}$ used for cooling the superconducting magnet, Penning traps, and the detection system is stabilized to $\delta z_{\text{lHe}} < 1$~mm/h along with the pressure of helium gas inside the magnet's bore to \hbox{$\delta p < 10 $~$\mu$bar/h}~\cite{reppPENTATRAPNovelCryogenic2012}. 

We employ the Fourier-transform ion-cyclotron-resonance detection technique~\cite{marshallFourierTransformIon1998} using cryogenic tank circuits connected to the Penning traps to pick up the small image current induced in the trap electrodes by the ion. The largest frequency $\omega_+$, and therefore the frequency with the highest contribution to the overall error, is measured using the phase-sensitive pulse and phase (PnP) method~\cite{cornellSingleionCyclotronResonance1989,cornellModeCouplingPenning1990}. This method, described in more detail below, sets an initial phase of the reduced cyclotron frequency, then the motion is left decoupled for a variable phase accumulation time $t_{\text{acc}}$ during which the phase can evolve freely, before reading out the final phase $\phi_{\text{meas}}$. The other two frequencies $\omega_z$ and $\omega_-$ are measured with the Fast-Fourier-Transform (FFT) dip and the double-dip technique, respectively~\cite{fengTankCircuitModel1996}.

\begin{figure}
    \centering
    \includegraphics{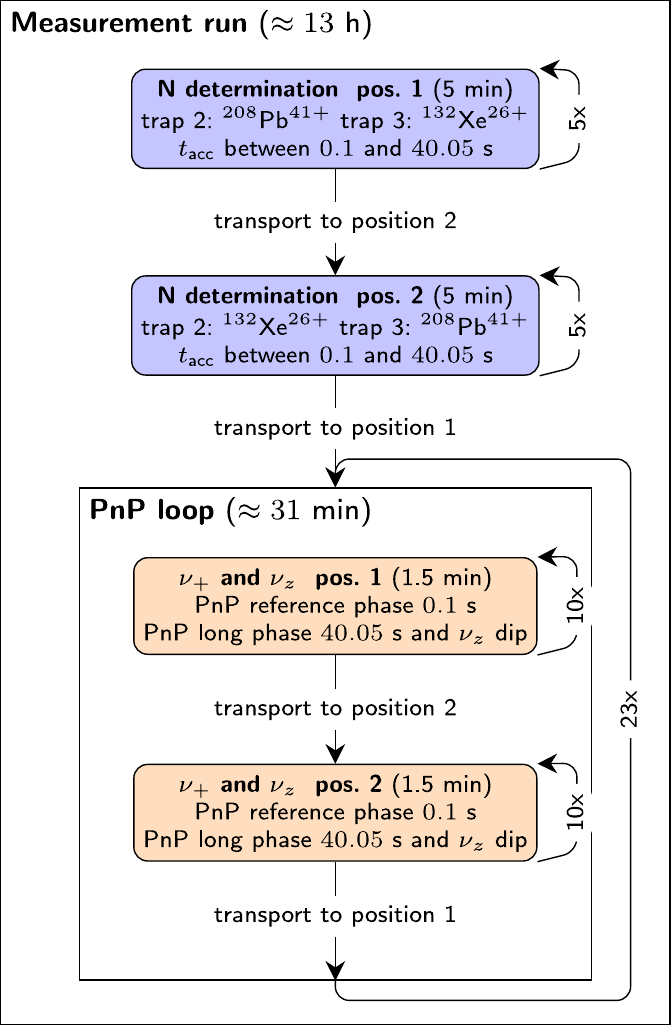}
  \caption{The figure depicts a flowchart of the measurement scheme and the relevant measurement times during the lead mass campaign. In blue at the top one can see the intial $N$ determination followed by the main measurement PnP loop in orange with the simultaneous measurement of axial frequency and modified cyclotron phase. The ions are frequently swapped in order to minimize the magnetic field drift between position 1 and 2.}
  \label{fig:measurementscheme}
 \end{figure}

The measurements of $^{208}$Pb$^{41+}$ versus $^{132}$Xe$^{26+}$ were carried out with the measurement scheme shown in \autoref{fig:measurementscheme}. After a rough estimate of all three frequencies of both ions in both positions, shown in \autoref{fig:Beamlines}b), using the dip and double-dip technique, the measurement run starts with an $N$ determination, with $N$ being the integer number of full turns of the reduced cyclotron motion during the phase accumulation time. This preparatory measurement is necessary before the actual PnP measurement because the cyclotron phase of the ion increases linearly with time with the increment of the frequency $\omega_+$ and will thereby pass a full turn of $2\pi$ many times during the phase accumulation time. The integer $N$ needs to be known in order to determine the modified cyclotron frequency through the total accumulated phase
\begin{equation}
    \phi(t_{\text{acc}}) = t_{\text{acc}} \times \omega_+ = 2\pi \times N + \phi_{\text{meas}} \text{ },
    \label{eq:phasedetermination}
\end{equation}
with $\phi_{\text{meas}}$ being the measured phase at the end of the accumulation time  $t_{\text{acc}}$. The $N$ determination utilizes 9 different phase accumulation times between 0.1 and 40.05~s and finds an $\omega_+$ for which each $N$ for all accumulation times is integer. 

A constant phase offset, unavoidable due to the phase readout, is cancelled out by subtracting a short reference phase with an accumulation time of 0.1~s from each long measurement phase. After concluding the $N$ determination in both traps for lead and xenon, the actual PnP loop is started, see \autoref{fig:measurementscheme} lower half. Here, a starting phase is imprinted on the modified cyclotron motion by an $\omega_+$ dipole excitation pulse. The phase can then evolve freely for $t_{\text{acc}}$ before the final modified cyclotron phase is imprinted on the axial motion by a coupling $\pi$-pulse, where it can be detected as an axial phase~\cite{cornellSingleionCyclotronResonance1989,cornellModeCouplingPenning1990}. All excitation and coupling pulses are shaped with a Tukey window~\cite{harrisUseWindowsHarmonic1978} in order to avoid systematic phase shifts during the phase readout. During the phase evolution time of the PnP sequence, an FFT axial-dip measurement is performed. This simultaneous phase determination and dip detection leads to a reduction of systematic effects associated with the temporal variation of the trap potential and the magnetic field, since they cancel out when calculating the free cyclotron frequency, using the invariance theorem in Eq. \eqref{eq:invariance}. After repeating the measurement of the two ions in trap 2 and 3 ten times, the ions are swapped. If $^{132}$Xe$^{26+}$ was in the trap, then $^{208}$Pb$^{41+}$ is swapped in and vice versa. This is repeated for around 12 hours before restarting the whole measurement scheme again with the $N$ determination. The magnetron frequency, being a factor $\approx1,600$ smaller than the modified cyclotron frequency, does not need to be measured repeatedly since the double-dip determination during the preparation phase is sufficiently precise. After one measurement run a relative statistical uncertainty of $\approx 10^{-11}$ is reached. 

To determine a neutral mass of $^{208}$Pb from the ions' free cyclotron frequency ratio we need to include the mass of the missing electrons in combination with their binding energies (in the following we always refer to the absolute binding energies). We employ the \textit{ab initio} fully relativistic multiconfiguration Dirac--Hartree--Fock (MCDHF) method~\cite{Grant1970,Desclaux1971,GRASP2018} to compute the binding energies $E_{\text{Xe}}$ and $E_{\text{Pb}}$ of the outermost 26 and 41~electrons, respectively, in neutral Xe and Pb atoms.
First, for the case of Xe, the ionization energy of the outermost 8~electrons has been experimentally determined to be of 424.7(7)~eV~\cite{NIST_ASD}. Thus, one only needs to calculate the binding-energy difference between the ground states of Xe$^{26+}$ ([Ar]$3d^{10}$~${}^1S_0$) and Xe$^{8+}$ ([Kr]$4d^{10}$~${}^1S_0$). Similarly, since the ionization energy of the outermost 4~electrons in neutral Pb has been measured to be 96.719\,04(61)~eV~\cite{NIST_ASD}, only the corresponding binding-energy difference between the ground states of Pb$^{41+}$ ([Kr]$4d^{5}$~${}^4P_{5/2}$) and Pb$^{4+}$ ([Xe]$4f^{14}5d^{10}$~${}^1S_0$) needs to be determined theoretically. 
In the following, we use $E_{\textrm{Xe}^{08-26}}$ and $E_{\textrm{Pb}^{04-41}}$ to represent these two terms.
 
Within the MCDHF scheme, the many-electron atomic state function (ASF)
is constructed as a linear combination of configuration state functions (CSFs) with common total angular momentum ($J$), magnetic ($M$), and parity ($P$) quantum \mbox{numbers:} $|\Gamma P J M\rangle = \sum_{k} c_k |\gamma_k P J M\rangle$.
The CSFs $|\gamma_k P J M\rangle$ are given as $jj$-coupled Slater determinants of one-electron orbitals, and $\gamma_k$ summarizes all the information needed to fully define the CSF, i.e. the orbital occupation and coupling of single-electron angular momenta. $\Gamma$ collectively denotes \mbox{all} the $\gamma_k$ included in the representation of the ASF. The set of CSF basis is generated by the GRASP2018 code~\cite{GRASP2018} via single and double (SD) excitation of electrons from the reference configurations to high-lying virtual orbitals. After solving the self-consistent MCDHF equations for the radial wavefunctions and the mixing coefficients $c_k$, the relativistic configuration interaction (RCI) method is applied to account for the corrections arising from the quantum electrodynamic terms and Breit interactions.
We systematically expand the size of the basis set by adding and optimizing virtual orbitals layer by layer up to $n=10$ ($n$ is the principle quantum number), with the final correlation energies being derived by extrapolating to $n=\infty$. 

As the ground states of Xe$^{26+}$ and Xe$^{8+}$ are both in closed-shell configurations, the CSF basis sets used for the calculations can be generated by allowing SD excitations from all the core electrons starting with the $1s$ orbitals. This gives a contribution from the SD electron correlation energy of 25.6~eV. The contributions from the Breit interactions and QED effects are 4.0 and 0.5~eV, respectively. 

For the calculation of Pb$^{41+}$, however, due to its open $4d^5$ configuration, the number of CSFs for $J=5/2$ easily grows above 4 million even for the SD exchanges of the $4s^24p^64d^5$ electrons and becomes not tractable. Therefore, to access $E_{\textrm{Pb}^{04-41}}$, one needs to construct an ion chain in the calculation to reduce the errors. We first calculate the binding-energy difference between Pb$^{4+}$ and Pb$^{22+}$ via SD excitations from core electrons down to the $3p$ subshell, and then calculate the binding-energy difference between Pb$^{22+}$ and Pb$^{36+}$ by allowing SD excitations of all the core electrons. Finally, the connection between Pb$^{36+}$ and Pb$^{41+}$ is bridged over Pb$^{42+}$ via SD excitations from the $4s$ orbitals. 
In total, the SD electron correlation effects contribute 58.1~eV to $E_{\textrm{Pb}^{04-41}}$. The Breit interactions and QED terms give rise to corrections of 9.3 and $-0.6$~eV, respectively, to the binding energy difference. 

Until now, only the SD correlation energies are included in the calculations. Considering that the uncertainties in Breit and QED terms are small, the neglected higher-order correlations will account for the systematic errors. To estimate these errors, we make use of the accurate ionization data of the outermost 8 and 4 electrons in Xe and Pb, respectively. 
The estimations are based on three observations. First, as a self-consistent theory, the MCDHF always approaches the real ground-state energy from above. Thus, the MCDHF binding energy of a given ionic ground state is always smaller than its real value. Second, for a given element, the contributions from higher-order correlation terms scale with the number of bound electrons. Therefore, the differences between the MCDHF binding energy and its real value is more likely to be smaller in highly charged ions. Lastly, within the same isoelectronic sequence, the contributions from higher-order correlations are always smaller for highly charged ions. This is because, perturbatively, in the denominator of each perturbation term, the energy differences between atomic states in highly charged ions are much larger than those in lower charged ions. 
As a result, for closed-shell ions, the calculated binding-energy differences based on the SD excitations are always smaller than their real values, but their deviations become narrower when the charge states become larger. 

For the case of Xe, we find that the calculated binding-energy difference between Xe$^{8+}$ and Xe is 3.5~eV smaller than the experimentally measured value of 424.7(7)~eV, with the single-electron ionization energies of Xe, Xe$^{7+}$, and Xe$^{8+}$ being 0.32~eV, 0.22~eV and 0.10~eV, respectively, smaller than their experimental measurements. Though the deviations of the single-electron ionization energies for some open-shell ions between Xe$^{8+}$ and Xe$^{26+}$ may be larger than the 0.22~eV deviation of that in Xe$^{7+}$, one can still conservatively expect that the average deviation of the 16 electrons will not be larger than 0.22~eV. This indicates that the systematic shift of $E_{\textrm{Xe}^{08-26}}$ shall be within 4.0~eV. To cover the range between 0 to $4.0$~eV, one can add a systematic correction of 2.0(2.0)~eV, with both systematic shift and uncertainty being 2.0~eV. This leads to $E_{\textrm{Xe}^{08-26}}=8,546.5(2.0)$~eV and $E_{\textrm{Xe}}=8,971.2(2.1)$~eV. 

With a similar procedure, the ionization energy of Pb$^{2+}$ and Pb$^{3+}$ are found to be 1.26 and 0.95~eV smaller than their experimental measured values when SD excitation from the $3p$ subshells are considered. This indicates an average deviation of $<1.0$~eV for the single-electron ionization energies for Pb$^{4+}$ to Pb$^{22+}$, and a systematic correction of 9.0(9.0)~eV to the binding energy of the corresponding 18 electrons. For the ions between Pb$^{22+}$ and Pb$^{41+}$, since they are close to the isoelectronic systems of Xe ions, one can conservatively assume a $<0.22$~eV average deviation of the corresponding single-electron ionization energies. After the summation, we obtain $E_{\textrm{Pb}^{04-41}}=28,633.9(9.1)$~eV and $E_{\textrm{Pb}}=28,730.6(9.1)$~eV. 

\section{Results}

After calculating the free cyclotron frequencies during each PnP loop, the interpolation method~\cite{Door2021} is applied to calculate the frequency ratios, see \autoref{fig:resultspic}. This method uses two consecutive cyclotron frequencies of one trap in position 1 and interpolates them to the time the cyclotron frequency of the position 2 in the same trap was measured. Then the frequency ratio of the interpolated value of position 1 and the matching value of position 2 can be formed cancelling out in first order the magnetic field drift over time. The linear drift of the magnetic field is $\Delta B/B = -2.3 \times 10^{-10}$~/h. The impact of the non-linear drifts of the B-field was thoroughly investigated and found insignificant on the level of the achieved statistical uncertainty. The final measured ion frequency ratio is 

\begin{align}
    R_{\text{meas}}-1 &= \frac{\nu(^{208}\text{Pb}^{41+})}{\nu(^{132}\text{Xe}^{26+})}-1  \nonumber \\  &= 1.252\,194\,24(9)\times 10^{-4} ,
    \label{eq:measuredratio}
\end{align}
with a relative statistical uncertainty of $9\times 10^{-12}$.

{
\tikzset{
  >={Stealth[width=2mm,length=2mm]}
  }
\begin{figure}
    \centering
\includegraphics[]{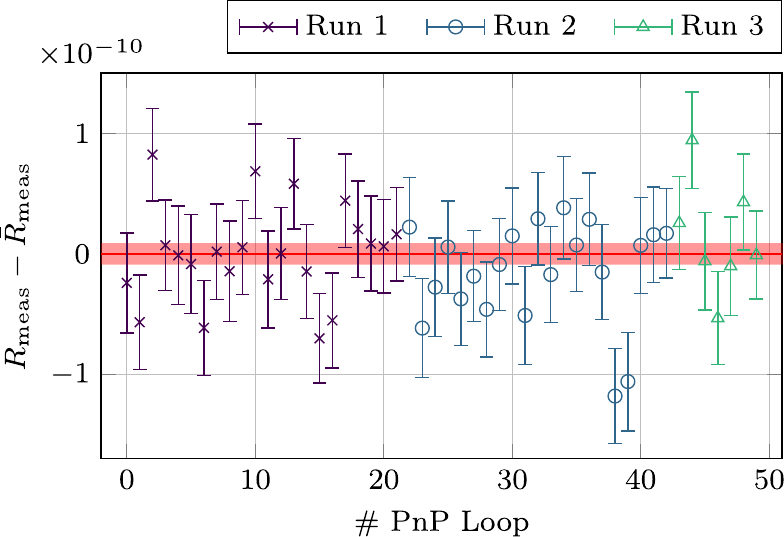}
\caption{The plot shows the frequency ratios for the different PnP loops. Each color represents a different measurement run. The $1 \sigma$ error band of the averaged ratio $\Bar{R}_{\text{meas}}$ is visualized in red.}
\label{fig:resultspic}
\end{figure}
}

The measured ratio is then corrected for known systematic shifts and their respective uncertainties. An overview of the relevant systematic effects and their size is listed in \autoref{tab:systematics}.
\begin{table}
\caption{\label{tab:systematics}
Systematic shifts and their uncertainties on the measured modified-cyclotron frequency ratio $R_{\text{meas}}$. For more details see text.}
\begin{tabularx}{\columnwidth}
{ccc}
\hline
Effect&Correction to $R_{\text{meas}}$&Uncertainty\\
&$(10^{-12})$&$(10^{-12})$\\
\specialrule{.1em}{.05em}{.05em} 
Image charge shift&185&9\\
Relativistic shift&0&4\\
\hline
Total&185&10\\
\hline
\end{tabularx}
\end{table}

The largest systematic shift comes from the image charge shift (ICS)~\cite{schuhImageChargeShift2019}. The highly charged ions induce an oscillating image charge in the trap electrodes. While this is necessary for detection, it causes a shift of the ions' frequencies by generating a counteracting electric field. The image charge shift depends strongly on the mass difference of the ions and on the radius of the trap, the latter being in the case of \pentatrap $5.000(5)$~mm. The ICS was determined to be $R_{\text{meas}}-\Tilde{R} = \Delta (R_{\text{meas}})_{\text{ICS}} = 1.85(9)\times 10^{-10}$, with $\Tilde{R}$ being the corrected ratio. In addition to this, the relativistic shift due to relativistic mass increase~\cite{brownGeoniumTheoryPhysics1986}
leads to another systematic uncertainty related to the size of the excited radii during the PnP measurement scheme: $\Delta (R_{\text{meas}})_{\text{rel}} = 0(4)\times 10^{-12}$. 

All other known systematic effects, due to e.g. trap potential anharmonicity, are on the order of $10^{-13}$ and below and are therefore neglected. Thus, the final $\omega_c$-ratio is $R-1 = 1.252\,192\,39 (9)(10)(13)\times 10^{-4}$, where the number in the first, second, and third brackets indicate the statistical, systematic, and total uncertainty, respectively. 

Combining the binding energies of the missing electrons calculated by theory, the experimentally determined mass ratio, and the mass excess of the reference isotope of $^{132}$Xe~\cite{redshaw2009improved,rischkaMassDifferenceMeasurementsHeavy2020} as listed in the AME2020, the mass excess of $^{208}$Pb is determined to be $-21,749.855(13)$~keV, which amounts to a neutral atomic mass of 207.976\,650\,571(14)~u. The new value improves the mass uncertainty of neutral $^{208}$Pb by a factor of 88 to a relative uncertainty of $\delta m/m=7\times 10^{-11}$ and shifts the mass excess value by $-1.4(1.1)$~keV. 

\section{Discussion and conclusion}
\begin{figure}
\includegraphics[]{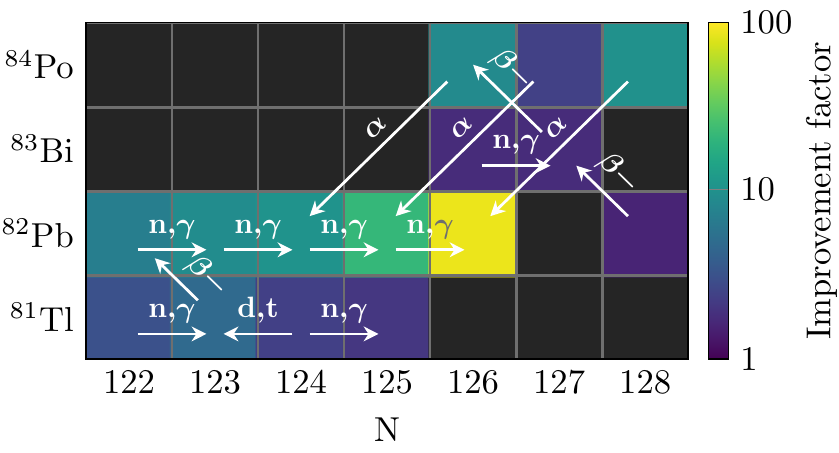}

    \caption{This colormap depicts a cutout of the nuclide chart. The color corresponds to the improvement factor (with 1 being no improvement) in mass precision after including the new mass value of $^{208}$Pb in the AME~\cite{wangAME2020Atomic2021}. The labeled arrows show the relevant connections for the mass determination of the different nuclides via different forms of decays from which the energy is known.}
    \label{fig:ImprovementPlot}
\end{figure}

In addition to the improvement of precision of the mass of $^{208}$Pb itself, our measurement also improves the masses of a series of lead isotopes, connected by \hbox{(n, $\gamma$)} reactions. So far, their mass precision was limited by the precision of $^{208}$Pb, but is now limited by the precision of the energy of the \hbox{(n, $\gamma$)} reactions. Furthermore, \autoref{fig:ImprovementPlot} shows the improvement in precision of, in total, 14 neighbouring nuclides' masses connected to the mass of $^{208}$Pb via different decays, e.g. $\alpha$ decay, for which the energy is well known. Since the new value of the mass excess of $^{208}$Pb is shifted downward, all these connected nuclides will be shifted down by $1.4$~keV. \autoref{tab:isotopeimprovement} lists the new mass values for all nuclides which were significantly improved. With the reported measurement we have established a new region in the nuclear chart with reference masses for experiments on heavy and superheavy nuclides. When measuring masses around $m=200$~u the error due to the reference mass will be as low as a few $10^{-10}$ and therefore negligible for mass determinations on radionuclides. 

Furthermore, with the new mass precision of $^{208}$Pb of $7 \times 10^{-11}$ the $g$-factor of the bound electron can be determined to the same level of precision. It therefore allows to carry out the experiments on $^{208}$Pb$^{81+}$ at Alphatrap and Artemis without having a large mass dependent error limiting their $g$-factor determination.

\begin{table*}
\addtolength{\tabcolsep}{-1pt}
\caption{\label{tab:isotopeimprovement}
New mass values of affected nuclides, when including the new mass value of $^{208}$Pb in the AME2020~\cite{wangAME2020Atomic2021}. }
\begin{tabular*}{\textwidth}{@{} l @{\extracolsep{\fill}} *{14}{c} @{}}
\hline
Z&A&el.&T$_{1/2}$~\cite{IAEANuclearData}&\multicolumn{2}{c}{AME2020 mass ($\mu$u)} &\multicolumn{2}{c}{Improved mass ($\mu$u)}& Improvement\\
&&&&&&&& factor\\
\specialrule{.1em}{.05em}{.05em} 

81 & 203 & Tl& stable &202,972,344.1& (1.3)& 202,972,342.7&(0.4)&3.2\\
81 & 204 & Tl& 3.783(12)~y &203,973,863.4& (1.2) & 203,973,862.01&(0.26)&4.8\\
82 & 204 & Pb& $1.4(6) \times 10^{17}$~y &203,973,043.5 & (1.2) & 203,973,042.09&(0.18)&7.0\\
81 & 205 & Tl& stable &204,974,427.3 & (1.3) & 204,974,425.9&(0.6)&2.4\\
82 & 205 & Pb& $1.70(9) \times 10^{7}$~y & 204,974,481.7& (1.2) & 204,974,480.26&(0.13)&9.2\\
81 & 206 & Tl&4.202(11)~min & 205,976,110.1 & (1.4) &205,976,108.7&(0.7)&2.1\\
82 & 206 & Pb& stable & 205,974,465.2 & (1.2) & 205,974,463.79&(0.12)&10.6\\
82 & 207 & Pb& stable &  206,975,896.8 & (1.2)& 206,975,895.39&(0.06)&21.6\\
82 & 208 & Pb& stable & 207,976,652.0 & (1.2) & 207,976,650.571&(0.014)&88.0\\
83 & 209 & Bi& $2.01(8) \times 10^{19}$~y & 208,980,398.6 & (1.5) &208,980,397.2&(0.8)&1.8\\
82 & 210 & Pb& 22.20(22)~y & 209,984,188.4 & (1.6) & 209,984,187&(1.0)&1.6\\
83 & 210 & Bi& 5.012(5)~d & 209,984,120.2 & (1.5) & 209,984,118.9&(0.8)&1.8\\
84 & 210 & Po& 138.376(2)~d& 209,982,873.7 & (1.2) & 209,982,872.27&(0.14)&8.8\\
84 & 211 & Po& 0.516(3)~s & 210,986,653.2 & (1.3) & 210,986,651.7&(0.6)&2.4\\
84 & 212 & Po& $294.3(8)$~ns & 211,988,868.0 & (1.2) & 211,988,866.55&(0.12)&10.1\\
\hline

\end{tabular*}
\end{table*}

\vspace{0.5cm}
\noindent
\textbf{Acknowledgments} This work is supported by the Max-Planck-Gesellschaft and the International Max-Planck Research School for Precision Tests of Fundamental Symmetries, the German Research Foundation (DFG) Collaborative Research Centre ``SFB 1225 (ISOQUANT)''. The project received funding from the European Research Council (ERC) under the European Union’s Horizon 2020 research and innovation programme under grant agreement number 832848 - FunI. Furthermore, we acknowledge funding and support by the Max Planck, RIKEN, PTB Center for Time, Constants and Fundamental Symmetries.

\bibliographystyle{spphys} 
\bibliography{science}

\begin{thebibliography}{10}
\providecommand{\url}[1]{{#1}}
\providecommand{\urlprefix}{URL }
\expandafter\ifx\csname urlstyle\endcsname\relax
  \providecommand{\doi}[1]{DOI \discretionary{}{}{}#1}\else
  \providecommand{\doi}{DOI \discretionary{}{}{}\begingroup
  \urlstyle{rm}\Url}\fi

\bibitem{oganessianSynthesisIsotopesElements2006}
Y.T. Oganessian, V.K. Utyonkov, Y.V. Lobanov, F.S. Abdullin, A.N. Polyakov,
  R.N. Sagaidak, I.V. Shirokovsky, Y.S. Tsyganov, A.A. Voinov, G.G. Gulbekian,
  S.L. Bogomolov, B.N. Gikal, A.N. Mezentsev, S.~Iliev, V.G. Subbotin, A.M.
  Sukhov, K.~Subotic, V.I. Zagrebaev, G.K. Vostokin, M.G. Itkis, K.J. Moody,
  J.B. Patin, D.A. Shaughnessy, M.A. Stoyer, N.J. Stoyer, P.A. Wilk, J.M.
  Kenneally, J.H. Landrum, J.F. Wild, R.W. Lougheed, Phys. Rev. C
  \textbf{74}(4), 044602 (2006).
\newblock \doi{10.1103/PhysRevC.74.044602}

\bibitem{blockDirectMassMeasurements2010}
M.~Block, D.~Ackermann, K.~Blaum, C.~Droese, M.~Dworschak, S.~Eliseev,
  T.~Fleckenstein, E.~Haettner, F.~Herfurth, F.P. He{\ss}berger, S.~Hofmann,
  J.~Ketelaer, J.~Ketter, H.J. Kluge, G.~Marx, M.~Mazzocco, Y.N. Novikov, W.R.
  Pla{\ss}, A.~Popeko, S.~Rahaman, D.~Rodr{\'i}guez, C.~Scheidenberger,
  L.~Schweikhard, P.G. Thirolf, G.K. Vorobyev, C.~Weber, Nature
  \textbf{463}(7282), 785 (2010).
\newblock \doi{10.1038/nature08774}

\bibitem{dworschakPenningTrapMass2010}
M.~Dworschak, M.~Block, D.~Ackermann, G.~Audi, K.~Blaum, C.~Droese, S.~Eliseev,
  T.~Fleckenstein, E.~Haettner, F.~Herfurth, F.P. He{\ss}berger, S.~Hofmann,
  J.~Ketelaer, J.~Ketter, H.J. Kluge, G.~Marx, M.~Mazzocco, Y.N. Novikov, W.R.
  Pla{\ss}, A.~Popeko, S.~Rahaman, D.~Rodr{\'i}guez, C.~Scheidenberger,
  L.~Schweikhard, P.G. Thirolf, G.K. Vorobyev, M.~Wang, C.~Weber, Phys. Rev. C
  \textbf{81}(6), 064312 (2010).
\newblock \doi{10.1103/PhysRevC.81.064312}

\bibitem{ramirezDirectMappingNuclear2012}
E.M. Ramirez, D.~Ackermann, K.~Blaum, M.~Block, C.~Droese, C.E. D{\"u}llmann,
  M.~Dworschak, M.~Eibach, S.~Eliseev, E.~Haettner, F.~Herfurth, F.P.
  He{\ss}berger, S.~Hofmann, J.~Ketelaer, G.~Marx, M.~Mazzocco, D.~Nesterenko,
  Y.N. Novikov, W.R. Pla{\ss}, D.~Rodr{\'i}guez, C.~Scheidenberger,
  L.~Schweikhard, P.G. Thirolf, C.~Weber, Science \textbf{337}(6099), 1207
  (2012).
\newblock \doi{10.1126/science.1225636}

\bibitem{eibachDirectHighprecisionMass2014}
M.~Eibach, T.~Beyer, K.~Blaum, M.~Block, C.E. D{\"u}llmann, K.~Eberhardt,
  J.~Grund, S.~Nagy, H.~Nitsche, W.~N{\"o}rtersh{\"a}user, D.~Renisch, K.P.
  Rykaczewski, F.~Schneider, C.~Smorra, J.~Vieten, M.~Wang, K.~Wendt, Phys.
  Rev. C \textbf{89}(6), 064318 (2014).
\newblock \doi{10.1103/PhysRevC.89.064318}

\bibitem{wangAME2020Atomic2021}
M.~Wang, W.J. Huang, F.G. Kondev, G.~Audi, S.~Naimi, Chinese Phys. C
  \textbf{45}(3), 030003 (2021).
\newblock \doi{10.1088/1674-1137/abddaf}

\bibitem{kellerbauerDirectAbsoluteMass2003}
A.~Kellerbauer, K.~Blaum, G.~Bollen, F.~Herfurth, H.J. Kluge, M.~Kuckein,
  E.~Sauvan, C.~Scheidenberger, L.~Schweikhard, Eur. Phys. J. D \textbf{22}(1),
  53 (2003).
\newblock \doi{10.1140/epjd/e2002-00222-0}

\bibitem{ketelaerTRIGASPECSetupMass2008}
J.~Ketelaer, J.~Kr{\"a}mer, D.~Beck, K.~Blaum, M.~Block, K.~Eberhardt,
  G.~Eitel, R.~Ferrer, C.~Geppert, S.~George, F.~Herfurth, J.~Ketter, S.~Nagy,
  D.~Neidherr, R.~Neugart, W.~N{\"o}rtersh{\"a}user, J.~Repp, C.~Smorra,
  N.~Trautmann, C.~Weber, Nuclear Instruments and Methods in Physics Research
  Section A: Accelerators, Spectrometers, Detectors and Associated Equipment
  \textbf{594}(2), 162 (2008).
\newblock \doi{10.1016/j.nima.2008.06.023}

\bibitem{grundFirstOnlineOperation2020}
J.~Grund, M.~Asai, K.~Blaum, M.~Block, S.~Chenmarev, C.E. D{\"u}llmann,
  K.~Eberhardt, S.~Lohse, Y.~Nagame, S.~Nagy, P.~Naubereit, J.J.W. {van de
  Laar}, F.~Schneider, T.K. Sato, N.~Sato, D.~Simonovski, K.~Tsukada, K.~Wendt,
  Nuclear Instruments and Methods in Physics Research Section A: Accelerators,
  Spectrometers, Detectors and Associated Equipment \textbf{972}, 164013
  (2020).
\newblock \doi{10.1016/j.nima.2020.164013}

\bibitem{sturmHighprecisionMeasurementAtomic2014}
S.~Sturm, F.~K{\"o}hler, J.~Zatorski, A.~Wagner, Z.~Harman, G.~Werth, W.~Quint,
  C.H. Keitel, K.~Blaum, Nature \textbf{506}(7489), 467 (2014).
\newblock \doi{10.1038/nature13026}

\bibitem{vogelElectronMagneticMoment2019}
M.~Vogel, M.S. Ebrahimi, Z.~Guo, A.~Khodaparast, G.~Birkl, W.~Quint, Annalen
  der Physik \textbf{531}(5), 1800211 (2019).
\newblock \doi{10.1002/andp.201800211}

\bibitem{kohlerIsotopeDependenceZeeman2016}
F.~K{\"o}hler, K.~Blaum, M.~Block, S.~Chenmarev, S.~Eliseev, D.A. Glazov,
  M.~Goncharov, J.~Hou, A.~Kracke, D.A. Nesterenko, Y.N. Novikov, W.~Quint,
  E.~Minaya~Ramirez, V.M. Shabaev, S.~Sturm, A.V. Volotka, G.~Werth, Nat Commun
  \textbf{7}(1), 10246 (2016).
\newblock \doi{10.1038/ncomms10246}

\bibitem{redshaw2009improved}
M.~Redshaw, B.J. Mount, E.G. Myers, Physical Review A \textbf{79}(1), 012506
  (2009)

\bibitem{rischkaMassDifferenceMeasurementsHeavy2020}
A.~Rischka, H.~Cakir, M.~Door, P.~Filianin, Z.~Harman, W.J. Huang,
  P.~Indelicato, C.H. Keitel, C.M. K{\"o}nig, K.~Kromer, M.~M{\"u}ller, Y.N.
  Novikov, R.X. Sch{\"u}ssler, C.~Schweiger, S.~Eliseev, K.~Blaum, Phys. Rev.
  Lett. \textbf{124}(11), 113001 (2020).
\newblock \doi{10.1103/PhysRevLett.124.113001}

\bibitem{reppPENTATRAPNovelCryogenic2012}
J.~Repp, C.~B{\"o}hm, J.R. {Crespo L{\'o}pez-Urrutia}, A.~D{\"o}rr, S.~Eliseev,
  S.~George, M.~Goncharov, Y.N. Novikov, C.~Roux, S.~Sturm, S.~Ulmer, K.~Blaum,
  Appl. Phys. B \textbf{107}(4), 983 (2012).
\newblock \doi{10.1007/s00340-011-4823-6}

\bibitem{rouxTrapDesignPENTATRAP2012}
C.~Roux, C.~B{\"o}hm, A.~D{\"o}rr, S.~Eliseev, S.~George, M.~Goncharov, Y.N.
  Novikov, J.~Repp, S.~Sturm, S.~Ulmer, K.~Blaum, Appl. Phys. B
  \textbf{107}(4), 997 (2012).
\newblock \doi{10.1007/s00340-011-4825-4}

\bibitem{zatorskiExtractionElectronMass2017}
J.~Zatorski, B.~Sikora, S.G. Karshenboim, S.~Sturm, F.~{K{\"o}hler-Langes},
  K.~Blaum, C.H. Keitel, Z.~Harman, Phys. Rev. A \textbf{96}(1), 012502 (2017).
\newblock \doi{10.1103/PhysRevA.96.012502}

\bibitem{brownPrecisionSpectroscopyCharged1982}
L.S. Brown, G.~Gabrielse, Phys. Rev. A \textbf{25}(4), 2423 (1982).
\newblock \doi{10.1103/PhysRevA.25.2423}

\bibitem{ovsyannikovFirstInvestigationsWarm1999}
V.P. Ovsyannikov, G.~Zschornack, Review of Scientific Instruments
  \textbf{70}(6), 2646 (1999).
\newblock \doi{10.1063/1.1149822}

\bibitem{StartseiteDREEBITGmbH}
Startseite {{DREEBIT GmbH}} - {{DREEBIT Company}} {DE}.
\newblock https://www.dreebit.com/de/

\bibitem{mickeHeidelbergCompactElectron2018}
P.~Micke, S.~K{\"u}hn, L.~Buchauer, J.R. Harries, T.M. B{\"u}cking, K.~Blaum,
  A.~Cieluch, A.~Egl, D.~Hollain, S.~Kraemer, T.~Pfeifer, P.O. Schmidt, R.X.
  Sch{\"u}ssler, C.~Schweiger, T.~St{\"o}hlker, S.~Sturm, R.N. Wolf,
  S.~Bernitt, J.R. {Crespo L{\'o}pez-Urrutia}, Review of Scientific Instruments
  \textbf{89}(6), 063109 (2018).
\newblock \doi{10.1063/1.5026961}

\bibitem{schweigerProductionHighlyCharged2019}
C.~Schweiger, C.M. K{\"o}nig, J.R. {Crespo L{\'o}pez-Urrutia}, M.~Door,
  H.~Dorrer, C.E. D{\"u}llmann, S.~Eliseev, P.~Filianin, W.~Huang, K.~Kromer,
  P.~Micke, M.~M{\"u}ller, D.~Renisch, A.~Rischka, R.X. Sch{\"u}ssler,
  K.~Blaum, Review of Scientific Instruments \textbf{90}(12), 123201 (2019).
\newblock \doi{10.1063/1.5128331}

\bibitem{Schweiger2021}
C.~Schweiger, et~al., in preparation  (2021)

\bibitem{bradburyAbsoluteValuesElectron1936}
N.E. Bradbury, R.A. Nielsen, Phys. Rev. \textbf{49}(5), 388 (1936).
\newblock \doi{10.1103/PhysRev.49.388}

\bibitem{marshallFourierTransformIon1998}
A.G. Marshall, C.L. Hendrickson, G.S. Jackson, Mass Spectrometry Reviews
  \textbf{17}(1), 1 (1998)

\bibitem{cornellSingleionCyclotronResonance1989}
E.A. Cornell, R.M. Weisskoff, K.R. Boyce, R.W. Flanagan, G.P. Lafyatis, D.E.
  Pritchard, Phys. Rev. Lett. \textbf{63}(16), 1674 (1989).
\newblock \doi{10.1103/PhysRevLett.63.1674}

\bibitem{cornellModeCouplingPenning1990}
E.A. Cornell, R.M. Weisskoff, K.R. Boyce, D.E. Pritchard, Phys. Rev. A
  \textbf{41}(1), 312 (1990).
\newblock \doi{10.1103/PhysRevA.41.312}

\bibitem{fengTankCircuitModel1996}
X.~Feng, M.~Charlton, M.~Holzscheiter, R.A. Lewis, Y.~Yamazaki, Journal of
  Applied Physics \textbf{79}(1), 8 (1996).
\newblock \doi{10.1063/1.360947}

\bibitem{harrisUseWindowsHarmonic1978}
F.~Harris, Proc. IEEE \textbf{66}(1), 51 (1978).
\newblock \doi{10.1109/PROC.1978.10837}

\bibitem{Grant1970}
I.P. Grant, Adv.~Phys. \textbf{19}, 747 (1970)

\bibitem{Desclaux1971}
J.P. Desclaux, D.F. Mayers, F.~O'Brien, J.~Phys.~B \textbf{4}, 631 (1971)

\bibitem{GRASP2018}
C.F. Fischer, G.~Gaigalas, P.~J{\"o}nsson, J.~Biero{\'n}, Comput. Phys. Commun.
  \textbf{237}, 184  (2019).
\newblock \doi{https://doi.org/10.1016/j.cpc.2018.10.032}

\bibitem{NIST_ASD}
A.~Kramida, {Yu.~Ralchenko}, J.~Reader, {and NIST ASD Team}.
\newblock {NIST Atomic Spectra Database (ver. 5.8), [Online]. Available:
  {\tt{https://physics.nist.gov/asd}} [2021, August 11]. National Institute of
  Standards and Technology, Gaithersburg, MD.} (2020)

\bibitem{Door2021}
M.~Door, et~al., in preparation  (2021)

\bibitem{schuhImageChargeShift2019}
M.~Schuh, F.~Hei{\ss}e, T.~Eronen, J.~Ketter, F.~{K{\"o}hler-Langes}, S.~Rau,
  T.~Segal, W.~Quint, S.~Sturm, K.~Blaum, Phys. Rev. A \textbf{100}(2), 023411
  (2019).
\newblock \doi{10.1103/PhysRevA.100.023411}

\bibitem{brownGeoniumTheoryPhysics1986}
L.S. Brown, G.~Gabrielse, Rev. Mod. Phys. \textbf{58}(1), 233 (1986).
\newblock \doi{10.1103/RevModPhys.58.233}

\bibitem{IAEANuclearData}
{{IAEA Nuclear Data Services}}.
\newblock https://www-nds.iaea.org/

\end{thebibliography}
\end{document}